\documentclass[10pt]{article}
\usepackage[colorlinks]{hyperref}
\usepackage{graphicx}

\RequirePackage{geometry}
\newcommand{\FILe}[1]{{\sf #1}}

\begin{document}

\title{QCDGPU: open-source package for Monte Carlo lattice simulations\\ on OpenCL-compatible multi-GPU systems}
\author{Vadim Demchik\thanks{\tt E-mail: vadimdi@yahoo.com},
~~~ Natalia Kolomoyets\thanks{\tt E-mail: rknv7@mail.ru}
~\\~\\~ {\small \sl{Dnipropetrovsk National University, 49010 Dnipropetrovsk, Ukraine}}} \maketitle

\begin{abstract}
The multi-GPU open-source package QCDGPU for lattice Monte Carlo
simulations of pure SU(N) gluodynamics in external magnetic field at finite temperature
and O(N) model is developed. The code is implemented in OpenCL, tested on AMD and
NVIDIA GPUs, AMD and Intel CPUs and may run on other OpenCL-compatible devices.
The package contains minimal external library dependencies and is
OS platform-independent. It is optimized for heterogeneous computing
due to the possibility of dividing the lattice into non-equivalent parts
to hide the difference in performances of the devices used.
QCDGPU has client-server part for distributed
simulations. The package is designed to produce lattice gauge configurations
as well as to analyze previously generated ones. QCDGPU may be executed
in fault-tolerant mode. Monte Carlo procedure core is based on
PRNGCL library for pseudo-random numbers generation on OpenCL-compatible
devices, which contains several most popular pseudo-random number
generators.
\end{abstract}


\section{Introduction}
Nowadays graphics processing units (GPU) play a rather important role in high-performance computing (HPC).
The proportion of computing systems equipped with the special accelerators (GPUs, MIC, etc.) is growing
among supercomputers, which is reflected in the well-known TOP500 list of the most powerful supercomputing
systems of the world \cite{top500:2013}. The most popular programming languages for general-purpose computing
on GPU (GPGPU) are CUDA (for NVIDIA GPUs only) and Open Computing Language (OpenCL) \cite{OpenCL:2011}.
Currently more than 80\% of all scientific researches are using GPU accelerators performed with
CUDA \cite{hgpustat:2013}.

One of the main approaches to study high-energy physics (HEP) phenomena is the lattice Monte Carlo simulations.
In 1974 Kenneth G. Wilson proposed a formulation of quantum chromodynamics (QCD) on a space-time lattice
\cite{Wilson:1974sk} -- a lattice gauge theory (LGT), which allows to calculate infinite-dimensional path
integral with the procedure of computation of finite sums. LGT has many important features, in particular,
LGT makes it possible to study low-energy limit of QCD, which is not achievable by analytic methods.
In the limit of an infinite number of lattice sites and zero lattice spacing, LGT becomes an ordinary
quantum field theory (QFT). Numerical results, obtained by means of lattice approximation, depend on
number of lattice sites, the using of lattices with large size is preferable. Moreover, some phenomena
could be studied on big lattices only, because small lattices are not sensitive to such effects.
The use of large lattices puts special demands on computer systems on which the investigation is performed.
Thus, the need for high computing performance in addition to the existence of well parallelized algorithms
makes the application of GPUs for lattice simulations particularly important. Now HEP is one
of the main consumers of supercomputing facilities.

Major HEP research collaborations develop software environments to achieve their scientific goals (see below).
The standard practice is to incorporate into packages special utilities for data exchange among collaborations.
While software development programmers optimize their code according to the hardware available for collaboration.
So, due to severe competition among hardware manufacturers it is necessary to take into consideration the
cross-platform principles while constructing a new HEP package.

\begin{figure}[ht]
  \begin{center}
    \includegraphics[width=0.6\textwidth]{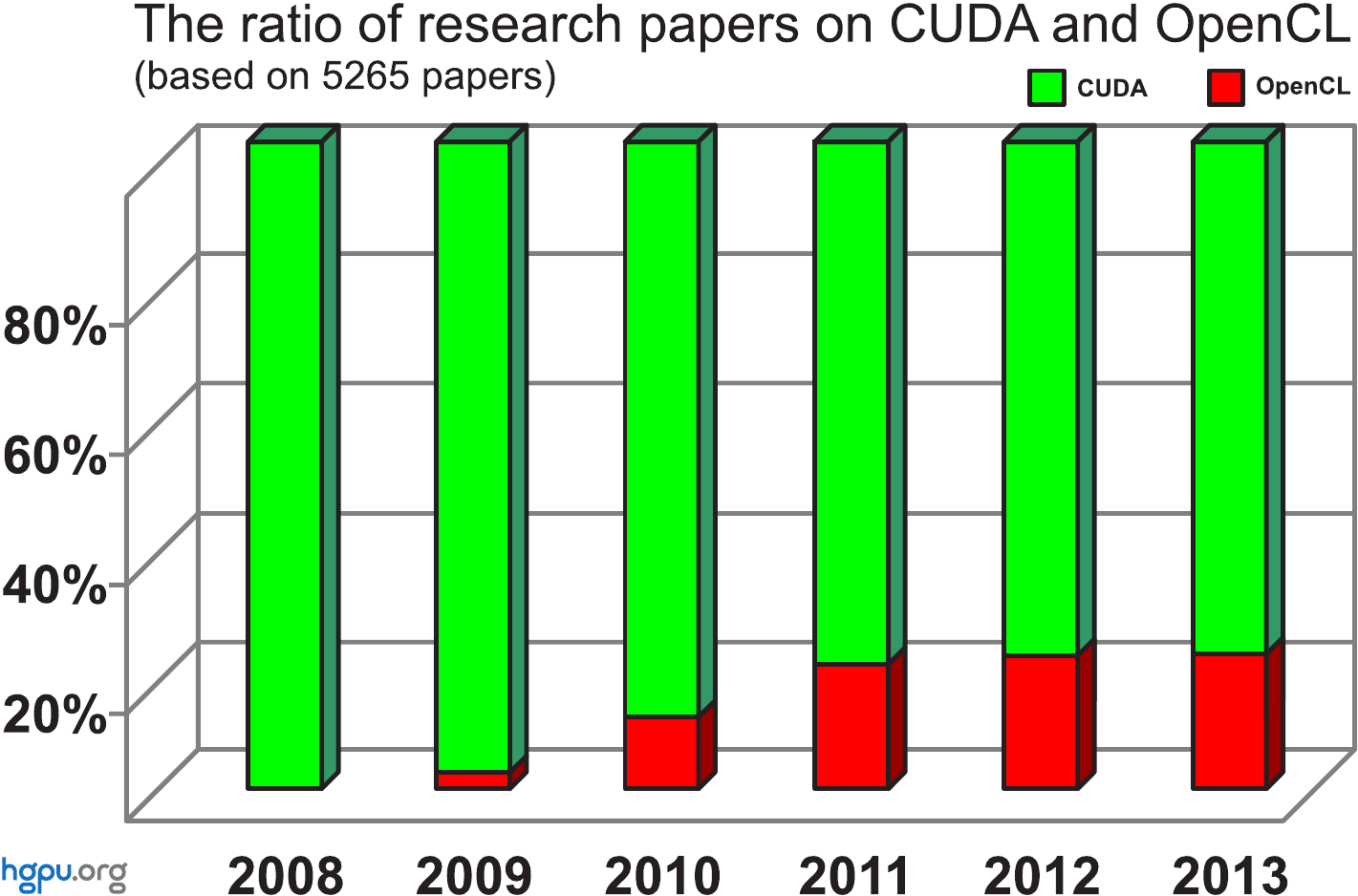}
    \caption{Currently $\sim 20\%$ of all scientific researches using GPU accelerators are performed with OpenCL \cite{hgpustat:2013}}
    \label{fig:openclstat}
  \end{center}
\end{figure}

Obviously, each software package has limited scope of scientific tasks, which could be solved with it.
Current HEP development leads to emerging the problems beyond this scope.
The spontaneous vacuum magnetization at high temperature \cite{Demchik:2012vf},
phase transition behavior in external fields \cite{Cea:2001pc}, dependence of the phase transition order in O(N)
models on coupling constant value \cite{Bordag:2012nh} and so on are some of such tasks.
In 2008 we created the IDS package \cite{Demchik:2009ni}, which allows to research quantum effects
in external chromomagnetic field. It was written in ATI IL for AMD/ATI GPUs and provided fast derivation
of a huge statistic, desired for solving applied tasks. After the deprecation to maintain ATI IL by AMD,
the demand to port the package to a modern GPGPU language has been appeared. Currently $\sim 20\%$ of all
scientific researches using GPU accelerators are performed with OpenCL \cite{hgpustat:2013} (see figure \ref{fig:openclstat}).
So, OpenCL was chosen to provide a multi-platform usage.

In this paper we present \FILe{QCDGPU} package and describe its program architecture and possibilities.
The general aim of the package is production of lattice gauge field configurations for pre described
models with following statistical processing of different measurable quantities. The current version of
\FILe{QCDGPU} allows to study SU(2), SU(3) gauge theories as well as O(N) models. The extension of available
groups in the package may be made by linking the file with appropriate algebra and core
of the Monte Carlo procedure for particular groups. The number of space-time dimensions is one of the run-time
parameters in \FILe{QCDGPU}. $3+1$ dimensional space-time is assumed by default. The list of measurable quantities
may be changed accordingly to the problem investigated. All lattice simulations as well as measurements
can be performed whether with single or double precisions. The package is easily-scalable to the number of available
devices.

The paper is organized as follows. The overview of existing packages is made in the Sect.\,\ref{sect:relworks}.
The structure of the \FILe{QCDGPU} package is shown in the Sect.\,\ref{sect:architecture}.
The description of multi-GPU mode and the capability of distributed simulations are provided
in the Sect.\,\ref{sect:multigpu}. Some performance results are shown in Sect.\,\ref{sect:performance}.
The last Sect.\,\ref{sect:discussion} is devoted to discussion of the scope of the package and summarizes the results.

\section{Related Works}
\label{sect:relworks}
While a great amount of new experimental data appeared, modern high-energy physics requires extremely
resource-intensive computations, in particular Monte Carlo lattice simulations. Therefore, only
big scientific collaborations, which have enough computation time on supercomputers can run such
simulations. As usual such collaborations (UKQCD, USQCD, TWQCD, PTQCD, etc.) have own software
packages for simulations. The most well-known among them are:
\begin{itemize}
  \item {\bf FermiQCD}: an open C++ library for development of parallel
Lattice Quantum Field Theory computations~\cite{DiPierro2005qx},
  \item {\bf MILC}: an open code of high performance research software
written in C for doing $SU(3)$ lattice gauge theory simulations on
several different (MIMD) parallel computers~\cite{MILC},
  \item {\bf QDP++/Chroma}: package supporting data-parallel programming constructs
for lattice field theory and in particular lattice QCD~\cite{Edwards:2004sx}.
\end{itemize}

The first paper relating application of GPUs in HEP lattice simulations was published in 2007 \cite{Egri:2006zm}.
The authors of this work used OpenGL as programming language and for the first time denoted
the need to store lattice data in the form of four-component vectors. Shortly after of the publication,
NVIDIA unveiled a new architecture CUDA and OpenGL ceased to be used as a GPGPU-computation language in further works.

Recently some open-source software packages have been developed targeted to use GPUs:
\begin{itemize}
  \item {\bf QUDA}: a library for performing calculations in lattice QCD on
CUDA-ready GPUs~\cite{Clark:2009wm},
  \item {\bf PTQCD}: a collection of lattice SU(N) gauge production programs on CUDA-ready GPUs~\cite{Cardoso:2011xu},
  \item {\bf cuLGT}: code for gauge fixing in lattice gauge field theories with CUDA-ready GPUs~\cite{Schrock:2013nta},
\end{itemize}
and several other packages with closed-access source codes (\cite{Bach:2012iw}, \cite{Chakrabarty:2012rv},
\cite{Chiu:2011bm}, \cite{Bonati:2011dv}, \cite{Alexandru:2011sc}, \cite{Kim:2010br}).
Some HEP collaborations link special GPU-libraries for their projects to engage
GPGPU computing possibility without code refactoring.
In particular, MILC collaboration uses QUDA package \cite{Shi:2010aq}, but only in single-device mode now.
USQCD collaboration also has powered its QDP++/Chroma software with CUDA \cite{Winter:2011an}.

Most of the GPU-targeted packages are based on NVIDIA CUDA environment. However, the development of HPC market
leads to increasing of cross-platform heterogeneous computing importance. In spite of unchallenged leadership
of CUDA the software packages adapted for CUDA-ready devices could not be executed on other (even hardware-compatible)
accelerators because of its closed standard.

\section{QCDGPU Package Architecture}
\label{sect:architecture}
The \FILe{QCDGPU} package architecture is based on the full platform-independence principle.
All modules of the package are written in C++ and OpenCL with minimal external
libraries dependence. \FILe{QCDGPU} can run equally well both on Windows operating system (OS),
and on Linux without any code modifications. The OS-independence is implemented by the
inclusion into the package of special file \FILe{platform.h}, which adapts the
OS-dependent commands by means of precompiler directives.

The package is tested on different OSs, OpenCL SDK of different vendors and on several devices:
\begin{itemize}
  \item {\bf OS:} Windows XP, Windows 7, OpenSUSE 11.4, OpenSUSE 12.2;
  \item {\bf SDK:} NVIDIA CUDA SDK 5.5, AMD APP SDK 2.8.1, Intel SDK for Applications OpenCL 2013;
  \item {\bf devices:} AMD Radeon HD7970, HD6970, HD5870, HD5850, HD4870, HD4850 (single-precision mode),
NVIDIA GeForce GTX 560 Ti, NVIDIA GeForce GTX 560 M, Intel Core i7-2600, Intel Core i7-2630QM, AMD Phenom II X6.
\end{itemize}
The package can be executed on all versions of OpenCL (1.0, 1.1, 1.2) without any code changes.

\subsection{Package Structure}

\begin{figure}[ht]
  \begin{center}
    \includegraphics[width=0.6\textwidth]{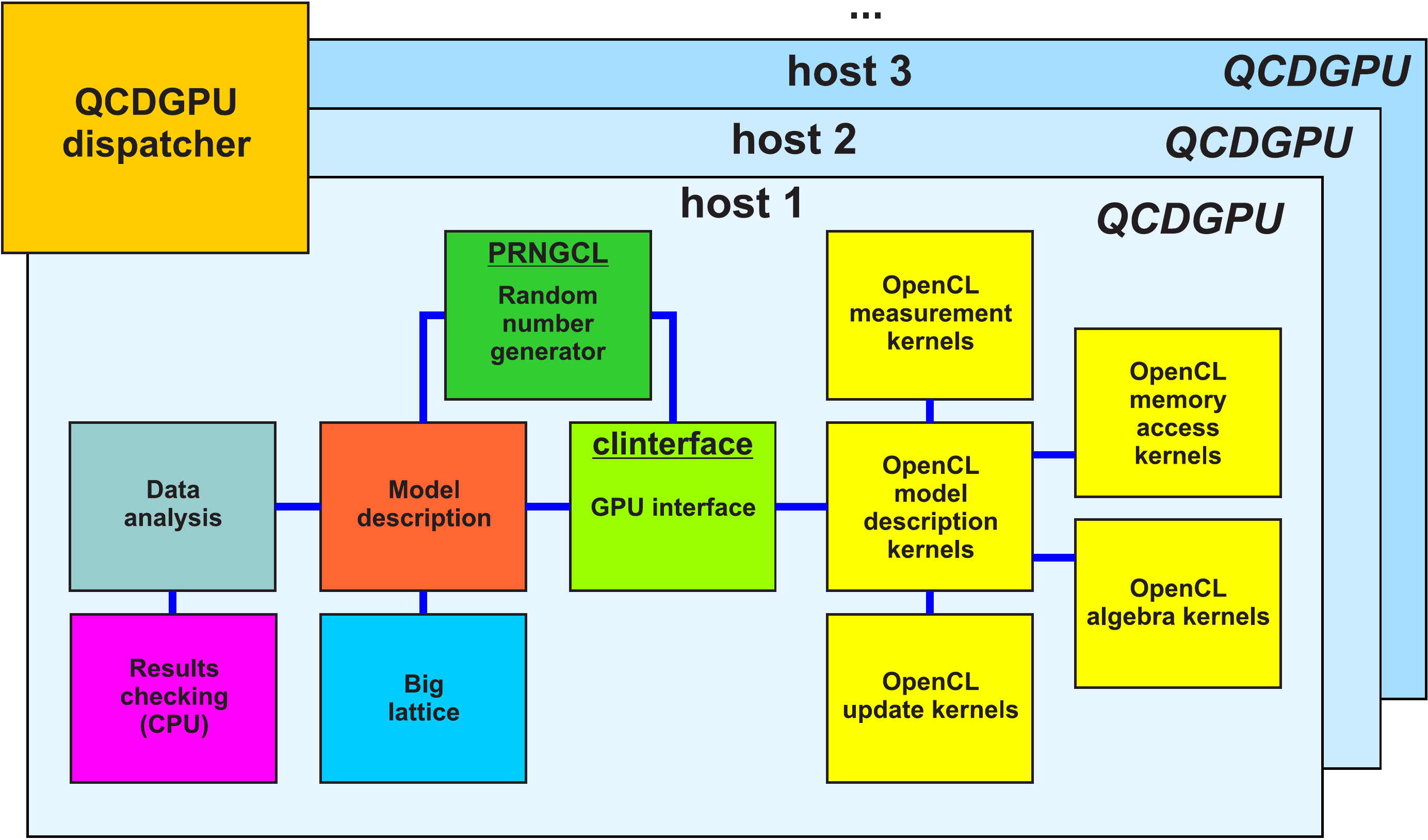}
    \caption{Structure of \FILe{QCDGPU} package.}
    \label{fig:structure}
  \end{center}
\end{figure}

The schematic diagram of the \FILe{QCDGPU} is shown in figure \ref{fig:structure}.
The package core compounds of the block \FILe{CLinterface}, which provides the
interaction of host code with computing devices. It performs all services for the
preparation of devices, run of kernels and release the host memory and devices.
Next important block is the block with physical model description \FILe{Model description}
(\FILe{SUNCL} or \FILe{ONCL} depending on the model under investigation). Memory organization
on the host in accordance with the physical model (gauge group, space-time dimension, etc.)
is made in this block, as well as preparation and configuration of kernels in accordance
with simulation conditions. Algorithms based on pseudo-random number generators are the
base of Monte Carlo procedure. Library \FILe{PRNGCL} performs the function of generating
pseudo-random numbers with required generator selection. The block \FILe{Big lattice} is designed
to provide the possibility to produce simulations on large lattices, and to use multiple
devices on a single host. The package performs all the necessary calculations on computing
devices and provides the final simulation results to the host. Statistical analysis of the results
over the run is performed by the block \FILe{Data analysis}. Validation of the data is
performed by the block \FILe{CPU Results checking}, which produces control measurements
of required quantities on the last gauge configuration by CPU means. The block is basically
designed for debugging (for example, if a device has non-ECC memory), it can independently
produce correspondent lattice gauge configurations on the same pseudo-random numbers as the
device. The block can be turned off to save host resources. The interaction of several copies of
the main program on different hosts is performed by a separate program \FILe{QCDGPU-dispatcher}.
The program realizes task scheduling for the available copies of the main program according
to the parameter space of the problem under investigation.

The results of simulations are written in separate text files that are sent for further
processing by external means. Each file contains the startup parameters needed for
reproduction of the run, as well as run average values and average configurations quantities
table. Also, for the possibility of resuming the interrupted simulation, the possibility
of regular saving of computation package state is realized, it allows to interrupt the simulation
at any time, and provides basic fault tolerance (power or hardware failure). Saving frequency
is set at the beginning of the simulations by the corresponding parameter. Saved file with
the computation state is portable (the calculation can be continued on another device).

Detailed description of the package blocks is below.

\subsection{CLinteface Module}
Block \FILe{CLinterface} is designed to hide all preparatory work for OpenCL-kernels startup
from the main program. At the same time fine adjustment of all programming units is available.
Every memory object and kernel obtain its own ID number and further kernel execution, memory object
binding to the kernel, results output and so on are carried out by this ID number. This principle
is initially used in the OpenCL standard \cite{OpenCL:2011}, but the numbering of objects is
produced mainly for the purposes of memory usage monitoring, users are not granted with the
possibility to use these internal ID numbers.

As far as lattice simulation implies multiply startup of the same  kernels, block \FILe{CLinterface}
allows to adjust parameters of each kernel startup by default. Meanwhile, there is only kernel's
ID number in startup command arguments, which makes the program code shorter. At the same time
there is a separate control of workgroupsize in case of Intel OpenCL SDK usage,
as it returns overestimated values workgroupsize for some devices.

The unit also controls compute device errors. All errors are recorded in .log-file.
Noncritical errors don't cause stop of the program.

The block also performs caching of previously compiled programs to reduce the startup time
of their launch. Built-in compute cache is realized only in the NVIDIA CUDA SDK. AMD APP SDK stores
only the last compiled program, while Intel SDK for applications OpenCL makes re-compilation
while each startup. In spite of compute cache in the NVIDIA CUDA SDK, there is a problem with the
re-compilation necessity of dependent files (included in the device-side OpenCL-program by directive
\FILe{\#include}) -- SDK up to 5.5 version does not monitor such file changes. All mentioned
above necessitates to create own compute cache. This compute cache is realized by creating
.bin-files with compiled code for a particular device with distinct compilation parameters.
.inf files are created along with these files, in which compilation-specific and
additional parameters (program number, platform, device, program options, MD5-hash of program
source, compilation timestamp) are indicated. For each source code MD5-hash is calculated, which is
de facto ID number of the source code version. New .bin and .inf are created if startup parameters
change. While changing MD5-hash of the source code, old .bin and .inf files are overwritten.
Such internal compute cache can be turned off in startup parameters.

The unit can perform a run-time profiling of kernels and memory objects. It allows to optimize
new kernels during designing of them. By default profiling is turned off.

Startup parameters are passed to kernels by 3 means:
\begin{enumerate}
  \item by determined parameters of internal precompiler;
  \item by constant buffers;
  \item directly by binding values.
\end{enumerate}
Undoubtedly, from the performance point of view, the most preferred way of passing parameters
to kernels is the first mean. But it necessitates its re-compilation. That's why only rarely
changed parameters are passed by this mean (lattice geometry, gauge group, precision and so on).
Often changed parameters are passed mainly by the second mean (coupling constant values, magnetic
field flux and so on), which are common for all kernels. Specific parameters for the particular
kernel are passed by the third mean (reducing size, memory offsets, etc.). Such division allows
to use computational time efficiently, which is formed from both program execution time,
and its compilation time.

\subsection{SUNCL and ONCL Modules}
The package block of \FILe{QCDGPU}, which is responsible for the physical model description, is modules
\FILe{SUNCL} and \FILe{ONCL}, which provide simulation SU(N) gluodynamics and O(N) models correspondingly.
As it is well known, the gauge fields in the group SU(N) are presented with $N \times N$ complex matrices.
These matrices are associated with lattice links. In case of O(N) models the fields are set with $N$-vectors,
which are associated with lattice sites. That's why it is possible to unify storage of lattice
data in the memory by the following means. The fastest index -- number of lattice site. Next index -- spatial
direction of lattice link. In case of O(N)-model, this index is not used. The slowest index is connected
with gauge group. At the same time group matrices are represented as structures of defined quantity
4-vectors, each of which contains a piece of information about corresponding matrix. This is due to the
architecture of GPU-devices memory.

In order to carry out lattice update, lattice is traditionally divided into even and odd sites
(checkerboard scheme), and, if necessary, into separate parts, which makes it possible to study
big lattices (see below). (Pseudo)heat-bath algorithm is used for SU(N) model update. Improved Metropolis
algorithm \cite{Bordag:2012nh} is used for O(N) update.

Due to the equalizing of array lengths, as well as offsets in accordance with workgroup size,
it achieves coalesced memory access, which has a positive effect on overall performance.

\subsection{Pseudo-Random Number Generators}
All pseudo-random numbers (PRN) needed for kernel operation are produced by own library \FILe{PRNGCL}.
The library is a porting and development of the library \FILe{PRNGCAL}, written for AMD/ATI GPUs on ATI CAL
\cite{Demchik:2010fd}. The most popular pseudo-random number generators (PRNG), which are used in HEP
lattice simulations (RANMAR, RANECU, XOR128, XOR7, MRG32k3a, RANLUX with different luxury levels), are
realized in it. Realization of Park-Miller PRNG and ``Constant generator'', which produces a given
constant, is included for testing and debugging purposes. The selection of the generator used is made by
one external parameter, which gives the possibility to check obtained results stability in relation to
PRNG used.

By ~default, ~the ~package ~generates ~PRNs ~by ~the ~number of threads equal to the parameter value
CL\_DEVICE\_IMAGE3D\_MAX\_WIDTH (maximal width of 3D image), which the device returns.
In practice this parameter connects to GPU device architecture and does not depend on the vendor.
Our observations show, that using this parameter as launched threads quantity allows
to achieve the best performance. Users can choose the quantity of threads for PRNs
producing manually.

As soon as almost all realized generators do not have a general scheme of multi-stream
execution (apart from MRG32k3a) PRNG parallelization is made by various
initializations of seed tables. Every PRNG thread keeps own seed table with own initial values.

The generator initialization is made by one value of the external parameter (\FILe{RANDSERIES}),
on which base PRNG seed tables are initialized. At the same time, if this parameter equals
zero, then system time is chosen as its value. The value of parameter \FILe{RANDSERIES} definitely
reproduces simulation results, so the value along with the name of PRNG used is presented
in output files.

\section{Multi-GPU Mode and Distributed Simulations}
\label{sect:multigpu}
One of the main features of package \FILe{QCDGPU} is its possibility to launch in multi-GPU mode.
At this time several devices on the host system can be used for one simulation. It is
evident that it is necessary to divide lattice into parts. In the general case dividing lattice
is made into unequal parts, which partly allows to hide the difference in devices performances.

Since the package is mainly designed to research finite-temperature effects,
dividing lattice into parts is made in first spatial coordinate direction $L_1$.
This is due to the fact that the temperature is associated with the time direction
$L_t$ and the case high temperature corresponds to $L_t<L_1$.

Divided lattice simulation differs from a full lattice simulation only in necessity
to carry out the exchanging values of boundary sites. To divide the lattice into parts
so-called second-level checkerboard scheme is used. It is carried out in the alternate
update of odd and even parts of the lattice. In this case, the border information
exchange is performed in asynchronous mode -- while even sites update is being made,
fulfil the information at border sites is taken place and vise verse. So while dividing
the lattice, it is preferable to use an even quantity of parts. Dividing the lattice is also
used in cases when compute device memory is not enough for the simulation.

The package also allows to make simulations on several hosts at the same time.
Each copy of the main computational program launches on the
corresponding host, while parameters passing and launch control are carried out
by external program \FILe{QCDGPU-dispatcher}. A very simple scheme is used for
interaction of the control and computational programs: in case of computational program
launch in this mode, the program waits for the special file \FILe{finish.txt} deleting
before simulation begins. If there is \FILe{QCDGPU} and \FILe{QCDGPU-dispatcher}
in general folder, it means ``the previous run is completed -- you can collect the
results, waiting for the next run''. After collecting files with simulation results,
the control program creates a special file \FILe{init.dat}, in which parameters of next
run are written, and file \FILe{finish.txt} is deleted. As soon as the computational
program does not find the file \FILe{finish.txt}, it reads new parameters of the launch
from the file \FILe{init.dat}. The cycle repeats. At the same time in case of using
several copies of computational program \FILe{QCDGPU-dispatcher} sequentially looks
through catalogues and set a new task for the first free host. Names of files with
results contain simulation finish time and unique prefix of computational program copy,
which makes each file unique.

\section{Performance Results}
\label{sect:performance}
In order to demonstrate some benchmarks we performed at several MC simulations for
O(4), SU(2), SU(3) models on various lattices on the following GPUs: NVIDIA GeForce GTX 560 M
(Windows 7), NVIDIA GeForce GTX 560 Ti, AMD Radeon HD 7970 (OpenSUSE 12.2), HD 6970, HD 5870
(OpenSUSE 11.4). For all MC simulations the ``hot'' lattice initialization and RANLUX
pseudo-random number generator with luxury level 3 were used. For O(4) model $NHIT=100$
tries were used to update each lattice site (this provides acceptance rate up to 50\%).
For SU(2) and SU(3) models $NHIT=10$ and reunitarization were used.
One bulk sweep was performed to decorrelate configurations to be measured. There are
two types of sweeps: thermalization and working sweeps.
During thermalization sweep each lattice element (sites or links) is updated. In working sweep
the lattice is updated and some quantities are measured. So, working sweeps are some longer than
thermalization sweeps. Here we present the timings for working sweeps only.

\begin{table}[ht]
  \begin{center}
    \caption{The timings of one working sweep for O(4), SU(2) and SU(3) models in
    single- and double-precision mode (in seconds).}
    \begin{tabular}{cccccc}
      \hline
      {\bf Model} & {\bf Device} & {\bf Parts} & {\bf Lattice} & {\bf Single} & {\bf Double} \\
      \hline
      O(4) & GTX560M  & 1 & $16^4$ & 0.08 & 0.12  \\
           &          &   & $24^4$ & 0.42 & 0.6   \\
           &          &   & $32^4$ & 1.31 & 1.87  \\
           &          & 4 & $48^4$ & 7.79 &11.11  \\
           & GTX560Ti & 1 & $32^4$ & 0.72 & 0.99  \\
           &          & 4 & $48^4$ & 4.64 & 6.10  \\
           & HD 5870  & 2 & $32^4$ & 0.96 & 1.56  \\
           & HD 6970  & 2 & $32^4$ & 0.73 & 1.3   \\
           & HD 7970  & 2 & $32^4$ & 0.66 & 0.95  \\
           &          &12 & $72^4$ &20.35 & 36.2  \\
           & HD 7970+GTX560Ti & 4 & $32^4$ & 0.72 & 0.96  \\
           &                  & 8 & $48^4$ & 2.97 & 4.1   \\
     SU(2) & GTX560Ti & 1 & $44^4$ & 0.95 & 1.14  \\
           & HD 5870  & 1 & $30^4$ & 0.22 & 0.24  \\
           & HD 6970  & 1 & $30^4$ & 0.20 & 0.22  \\
           & HD 7970  & 1 & $48^4$ & 4.64 & 6.10  \\
     SU(3) & GTX560Ti & 1 & $32^4$ & 0.56 & 1.17  \\
           &          &   & $36^4$ & 0.90 & 1.86  \\
           & HD 6970  & 1 & $24^4$ & 0.24 & 0.23  \\
           & HD 7970  & 1 & $28^4$ & 0.13 & 0.22  \\
      \hline
    \end{tabular}
    \label{tab:tab1}
  \end{center}
\end{table}

The performance results are collected in the Table 1. The gauge model and computing devices
used in MC simulations are shown in first and second columns, correspondingly. Due to the
memory limit of particular computing device whole lattice should be divided into several
parts to perform MC simulations. The number of parts is presented in the third column.
Obviously, the bigger number of lattice parts means the bigger number of boundary elements
transmission between host and device (or between devices in multi-GPU mode),
which reduces the overall performance. The last two columns contain the timings for single-
and double-precision mode simulations, correspondingly.

Among all the data in the table the case of cooperative simulation of one lattice on two
different OpenCL platforms is shown (AMD Radeon HD 7970 on AMD APP SDK 2.8 and NVIDIA GeForce
GTX 560 Ti on NVIDIA CUDA 5.5). In this case the timings are better only on 10-25\% than
simulation on a single best device. Nevertheless, simultaneous multi-platform simulations
might be interesting for very big lattices.

In \cite{Cardoso:2011xu} Cardoso and Bicudo reported the timings $6\times 10^{-4}$ ($7\times 10^{-4}$
with double precision) and $2\times 10^{-3}$ ($4\times 10^{-3}$ with DP) seconds per single sweep
on lattice $8^4$ for SU(2) and SU(3) models, respectively. The authors used NVIDIA GeForce GTX 580
and NVIDIA CUDA. We performed the same performance measurements with \FILe{QCDGPU} (lattice sweep
with reunitarization, without any measurements) and obtain the following values in seconds:
$6\times 10^{-4}$ ($10^{-3}$ with DP) for SU(2) and
$1.3\times 10^{-3}$ ($4.9\times 10^{-3}$ with DP) for SU(3) on NVIDIA GeForce GTX 560 Ti
and $2.0\times 10^{-3}$ ($2.6\times 10^{-3}$ with DP) for SU(2) and
$3.4\times 10^{-3}$ ($5.1\times 10^{-3}$ with DP) for SU(3) on AMD Radeon HD 7970.

In actual MC simulations the trivial parallelization scheme (each computing device receives
from dispatcher unique parameters set for simulation of whole lattice) we often use. The best
performance results are obtained in this case.
Undoubtedly, due to many tuning parameters (such as number of lattice parts, size of parts
for different computing devices, part sequence, workgroup sizes, etc.) the \FILe{QCDGPU} package
performance of multi-GPU mode is the subject for special research.

\section{Discussion}
\label{sect:discussion}
In the present work a new package \FILe{QCDGPU} is introduced. This package is designed
for Monte Carlo simulations of SU(N) gluodynamics in external field and O(N) model
on OpenCL-compatible devices. The package allows to carry out the simulations for very big lattices
in single- or multi-GPU mode. Simulations can be run with single or double precision.
The package claims low demands to the host CPU and practically does not load it, which provides
the possibility to use the package along with other traditional computational programs.

The \FILe{QCDGPU} package allows to investigate most popular LGT models in N-dimensional
space-time. The architecture of the \FILe{QCDGPU} provides a possibility to easily add
new LGT model into the package.

If the size of the lattice under investigation allows its location in device memory,
all necessary operations are carried out on device memory, the result returns to the host program
after finishing the simulation. If the lattice is too big to locate it in device memory, it is
divided into parts and work on separate parts is made by all computing devices available at
the host. \FILe{QCDGPU} package allows the simultaneous run of several instances of
the computational program on all tested OSs.

The current version of the program uses trivial parallelization scheme to distribute computing --
every computing node gets separate task for simulation. At the same time OS type on every used
host is not important. The main requirement is to create a folder with shared access at the host.
Now it is made within the local network framework and by the means of virtual private network
(VPN) organization for remote nodes. Using small pieces of text information between task scheduling
module \FILe{QCDGPU-dispatcher} and the host does not impose load on the network.

Built-in mechanism for saving the computational state while achieving some conditions (every N sweeps
or every M seconds) allows to interrupt long calculations without threat of data loss and to continue
them on another available device. It is also very useful while often power failures.

At present we are working on including to the package the following possibilities:
\begin{itemize}
  \item using the base of the built-in profiling mechanism, as well as additional micro-benchmarks to make
automatic adjustments of package start-up parameters;
  \item realization of RHMC algorithm for including fermionic fields on a lattice;
  \item running of one simulation on several hosts at the same time on the base of MPI, which is very
important in case of accounting of fermionic fields. Using mixed-precision possibility will be realized
to reduce amount of exchanging information.
\end{itemize}

In this work we did not bring attention to the details of realized algorithms, as well as to detailed
description of the package parameters. The package is being constantly developed. In the first place
methods and algorithms needed for actual research in the Quantum Chromoplasma Laboratory of Dnepropetrovsk
National University is realized. Obtained physical results are published in particular in
the works: \cite{Demchik:2012vf}, \cite{Bordag:2012nh}, etc.

Source codes of the package \FILe{QCDGPU} and some examples of result files are in the open
access at \\\url{https://github.com/vadimdi/QCDGPU}.

\end{document}